\RequirePackage{fix-cm}
\documentclass[twocolumn,natbib]{svjour3}          
\smartqed  
\usepackage{graphicx}
\usepackage{amsmath,amssymb}
\usepackage{color,xcolor}
\usepackage{enumerate}
\usepackage{setspace}
\usepackage{lastpage}
\usepackage{graphicx}
\usepackage{adjustbox}
\usepackage{float}
\usepackage{xspace}
\usepackage{booktabs}
\usepackage{url}
\usepackage[hidelinks]{hyperref}
\usepackage{numprint}
\npthousandsep{\,}
\usepackage{algpseudocode}
\usepackage{algorithm}

\usepackage{calc}
\usepackage{xstring}

\makeatletter
\newcommand*{\toccontents}{\@starttoc{toc}}
\makeatother

\DeclareGraphicsExtensions{.pdf,.png,.eps}

\newcommand{\eg}{e.\,g.}
\newcommand{\ie}{i.\,e.}

\let\originalleft\left
\let\originalright\right
\renewcommand{\left}{\mathopen{}\mathclose\bgroup\originalleft}
\renewcommand{\right}{\aftergroup\egroup\originalright}

\makeatletter

\let\oldnorm\|

\definecolor{orange}{rgb}{1,.647,0}

\renewcommand{\|}{|\!|}         
\newcommand{\T}{{}^{\mathsf{T}}}
\newcommand\code[1]{{\tt#1}}

\def\R{\code{R}}

\DeclareMathOperator{\E}{E}                       

\DeclareMathOperator{\e}{e}                       

\DeclareMathOperator{\CV}{CV}

\renewcommand{\d}{\mathsf{\,d}}

\newcommand{\cO}{{\cal{O}}}

\newcommand{\bbeta}{{\boldsymbol{\beta}}}

\newcommand{\bzeta}{{\boldsymbol{\zeta}}}
\newcommand{\btheta}{{\boldsymbol{\theta}}{}}

\newcommand{\bxi}{{\boldsymbol{\xi}}}

\newcommand{\bSigma}{{\boldsymbol{\Sigma}}{}}

\newcommand{\A}{{\mathbf{A}}}
\newcommand{\B}{{\mathbf{B}}}
\newcommand{\C}{{\mathbf{C}}}
\newcommand{\D}{{\mathbf{D}}}
\newcommand{\F}{{\mathbf{F}}}
\newcommand{\G}{{\mathbf{G}}}
\renewcommand{\H}{{\mathbf{H}}}
\newcommand{\I}{{\mathbf{I}}}
\newcommand{\J}{{\mathbf{J}}}
\newcommand{\K}{{\mathbf{K}}}
\renewcommand{\L}{{\mathbf{L}}}

\newcommand{\N}{{\mathbf{N}}}
\renewcommand{\O}{{\mathbf{O}}}
\renewcommand{\P}{{\mathbf{P}}}  
\newcommand{\Q}{{\mathbf{Q}}}

\renewcommand{\S}{{\mathbf{S}}}
\newcommand{\U}{{\mathbf{U}}}
\newcommand{\V}{{\mathbf{V}}}
\newcommand{\W}{{\mathbf{W}}}
\newcommand{\X}{{\mathbf{X}}}
\newcommand{\Y}{{\mathbf{Y}}}
\newcommand{\Z}{{\mathbf{Z}}}

\newcommand{\0}{{\mathbf{0}}}

\renewcommand{\a}{{\textbf{\textit{a}}}}
\renewcommand{\b}{{\textbf{\textit{b}}}}
\renewcommand{\c}{{\textbf{\textit{c}}}}
\newcommand{\f}{{\textbf{\textit{f}}}}
\newcommand{\g}{{\textbf{\textit{g}}}}
\newcommand{\h}{{\textbf{\textit{h}}}}

\renewcommand{\l}{{\textbf{\textit{l}}}}
\renewcommand{\k}{{\textbf{\textit{k}}}}
\newcommand{\m}{{\textbf{\textit{m}}}}

\newcommand{\p}{{\textbf{\textit{p}}}}
\newcommand{\q}{{\textbf{\textit{q}}}}
\renewcommand{\r}{{\textbf{\textit{r}}}}
\newcommand{\s}{{\textbf{\textit{s}}}}
\renewcommand{\t}{{\textbf{\textit{t}}}}
\renewcommand{\v}{{\textbf{\textit{v}}}}
\newcommand{\w}{{\textbf{\textit{w}}}}
\newcommand{\x}{{\textbf{\textit{x}}}}
\newcommand{\y}{{\textbf{\textit{y}}}}
\newcommand{\z}{{\textbf{\textit{z}}}}

\newcommand{\M}{{\textbf{\textit{M}}}}

\newcommand*{\stack@relbin}[3][]{%
  \mathop{#3}\limits
  \toks@{#1}%
  \edef\reserved@a{\the\toks@}%
  \ifx\reserved@a\@empty\else_{#1}\fi
  \toks@{#2}%
  \edef\reserved@a{\the\toks@}%
  \ifx\reserved@a\@empty\else^{#2}\fi
  \egroup
}%
\renewcommand*{\stackrel}{%
  \mathrel\bgroup\stack@relbin
}
\newcommand*{\stackbin}{%
  \mathbin\bgroup\stack@relbin
}

\journalname{}

\begin{document}

\title{Parallel cross-validation: a scalable fitting method for Gaussian process models\thanks{%
F.~Gerber is supported by the Swiss Nation Science Foundation (Fellowships P2ZHP2\_174828 and P400P2\_186680).
  }
}

\titlerunning{Parallel CV}        

\author{Florian Gerber         \and
        Douglas W.~Nychka  
}

\authorrunning{F.~Gerber, D.~W.~Nychka} 

\institute{F.~Gerber (corresponding author) \at
           Department of Applied Mathematics and Statistics, Colorado School of Mines, Golden CO, USA\\
           \email{gerber@mines.edu}, \href{https://orcid.org/0000-0001-8545-5263}{ORCID: 0000-0001-8545-5263}           
           \and
           D.~W.~Nychka \at
           Department of Applied Mathematics and Statistics, Colorado School of Mines, Golden CO, USA\\
           \email{nychka@mines.edu}, \href{https://orcid.org/0000-0003-1387-3356}{ORCID: 0000-0003-1387-3356},           
}

\date{December 30, 2019}

\maketitle

\begin{abstract}
Gaussian process (GP) models are widely used to analyze spatially referenced data and to predict values at locations without observations.
In contrast to many algorithmic procedures, GP models are based on a statistical framework, which enables uncertainty quantification of the model structure and predictions.
Both the evaluation of the likelihood and the prediction involve solving linear systems.
Hence, the computational costs are large and limit the amount of data that can be handled. 
While there are many approximation strategies that lower the computational cost of GP models,
they often provide only sub-optimal support for the parallel computing capabilities of current (high-performance) computing environments.
We aim at bridging this gap with a parameter estimation and prediction method that is designed to be parallelizable. 
More precisely, we divide the spatial domain into overlapping subsets and use cross-validation (CV) to estimate the covariance parameters in parallel.
We present simulation studies, which assess the accuracy of the parameter estimates and predictions.
Moreover, we show that our implementation has good weak and strong parallel scaling properties.
For illustration, we fit an exponential covariance model to a scientifically relevant canopy height dataset with $5$~million observations.
Using $512$ processor cores in parallel brings the evaluation time of one covariance parameter configuration to less than $1.5$~minutes.
The parallel CV method can be easily extended to include approximate likelihood methods, multivariate and spatio-temporal data, as well as non-stationary covariance models.  
\end{abstract}

\section{Introduction}
An important benefit of the rapid advances in computing, data storage, and remote sensing is the availability of large spatial and space-time datasets, which help to address substantial scientific questions.
Such data are relevant in weather and climate applications, but also contribute to a better understanding of processes on the Earths surface \citep{Hmim:etal:13}. 
For example, densely spaced Light Detection and Ranging (LiDAR) measurements from overflights of forested regions provide a unique opportunity to study forest ecology and monitor changes over time \citep{Lefs:etal:02}.
A hallmark of many of these datasets is the large numbers of often irregularly spaced spatial observations, which poses statistical as well as computational challenges and motives our study. 

It is important to base predictions from large spatial datasets on a sound statistical framework to provide reliable measures of uncertainty in the predictions and other model components.
This is in contrast to more algorithmic approaches that just focus on computationally efficient predictions \citep{Gerb:etal:18,Weis:etal:14},
A starting point for many statistical models is assuming an underlying GP for a field that represents the data directly \citep{Wikl:etal:19} or a latent field connected to the data \citep{Bane:etal:14}.
Although there exists a mature methodology for such models and their application to large datasets, the ever-increasing amount of data remains challenging and motivates current research \citep{Liu:etal:18,Heat:Gerb:etal:18}.  
However, only a few methods can exploit the capabilities of current high-performance computing infrastructures, and this study is an advance in that direction.

Global optimization of the likelihood, a Bayesian posterior, or a CV based loss function is known to provide statistically accurate estimates of model parameters.
But a computationally attractive alternative is to break up the spatial domain into sub-domains (\eg, tiles) and analyze each sub-domain separately. 
This approach reduces the computational workload and is amenable to parallelization.
However, a naive implementation leads to a model that substantially differs from a global model.
Major drawbacks are that borrowing strength for global statistical parameters is impossible and predictions near the boundaries of the sub-domains can be poor.
In this work we show that a subsetting approach featuring overlapping sub-domains can achieve results that are comparable to those from the corresponding global model.

We focus on out-of-sample CV for estimating covariance parameters \citep{Rasm:Chri:05}, as it is efficient relative to maximum likelihood (ML) and facilitates the combination of local goodness-of-fit measures into global ones.
Our approach takes advantage of the well-known \emph{screening effect} in spatial prediction \citep{Stei:02}, whereby conditioning on nearby observations decreases the statistical value of more distant ones.
Based on this effect a relatively small overlap of the subsets is sufficient for a good approximation of the global model. 
Although both the subsetting and CV ideas are not new, our combination is a flexible and scalable fitting method, which can take into account information from millions of locations and still accurately approximates the GP models commonly used for spatial data analysis.

\section{Method}\label{sec:method}
\subsection{Spatial Gaussian process model}\label{sec:model}
For the spatial location~$s$ in the domain $D \subset \mathbb{R}^2$ the process $Y(s)\in \mathbb{R}$ is a GP if 
all finite dimensional realizations $\y=(y_1, \dots y_n)\T$ of $Y(s)$ at the locations $\s=(s_1,\dots, s_n)\T$ follow a multivariate Gaussian distribution.
In the following we assume that a vector of spatial observations is distributed as
\begin{equation}\label{eq:mod}
  \begin{split}
    \y \sim \mathcal{N}\left(\0, \sigma^2 \bSigma(\btheta) + \tau \I  \right),
  \end{split}
\end{equation}%
where $\sigma^2>0$ is the marginal variance, $\tau \geq 0$ is the measurement error (or nugget effect), and $\btheta$ are parameters of the $n \times n$ covariance matrix~$\bSigma(\btheta)$ that is derived from a process covariance function $c(s_1,s_2, \btheta)$.
For this model two times the negative log-likelihood of $\bxi=(\sigma^2, \tau, \btheta\T)\T$ given $\y$ is
\begin{equation}\label{eq:lik}
  \begin{split}
    -2l(\bxi; \y)=&\,\, n\log(2\pi) + \log \det\left(\sigma^2\bSigma(\btheta) + \tau \I\right) \,+ \\
    &\,\,\y^\top\left(\sigma^2\bSigma(\btheta) + \tau \I\right)^{-1} \y,\\
  \end{split}
\end{equation}
and the corresponding ML estimate $\widehat \bxi_\text{ML}$ minimizes~\eqref{eq:lik} with respect to~$\bxi$.  

The GP model can be used to predict the values of $Y(s)$ at any location $s \in D$.
Given the observations $\y$ and the parameters $\bxi$ the best linear unbiased predictor of~$Y(s)$ is also known as the kriging predictor~\citep{Stei:99}.
To formalize, let $s_p$ be a spatial location in $D$ at which we would like to predict~$Y(s_p)$.
Then the simple kriging prediction of $\y_p$~is
\begin{equation}\label{eq:krig}
\begin{split}
  \widehat \y_{P,\bxi} =&\,\,\sigma^2 \bSigma_p(\btheta)\T\, \left(\sigma^2\bSigma(\btheta) + \tau \I\right)^{-1}\, \y\\
 =&\,\, \bSigma_p(\btheta)\T \,\left(\bSigma(\btheta) + \frac\tau{\sigma^2} \I\right)^{-1}\, \y,
\end{split}
\end{equation}
where $\sigma^2\bSigma_p(\btheta)$ is the $n \times 1$ cross-covariance matrix of $\y$ and~$Y(s_p)$.
Note that instead of both $\sigma^2$ and $\tau$ only the noise-to-signal ratio $\lambda=\tau/\sigma^2$ is relevant for the prediction, and we reduce the parameter vector to $\bzeta=(\lambda, \btheta\T)\T$ and write $\widehat \y_{P,\bzeta}$. 

Both the evaluation of the likelihood in~\eqref{eq:lik} and the prediction in~\eqref{eq:krig} require $\mathcal{O}\left(n^3\right)$ operations and $\mathcal{O}\left(n^2\right)$ memory.
Hence, the exact computations become intractable for large datasets, which motivates more efficient approximate methods.

\subsection{CV based covariance parameter estimation}
An alternative to ML estimation of the covariance parameters is CV~\citep{Rasm:Chri:05}.
The main idea is to divide~$\y$ into $n_T\in\mathbb{N}$ training data $\y_T$ as well as $n_V\in\mathbb{N}$ validation data~$\y_V$ and to assess the goodness of a parameter configuration, $\bzeta$, by how accurately $\widehat \y_V$ predicts~$\y_V$. 
Although different metrics can be used to measure the accuracy of the predictions~\citep{Zhan:Wang:10}, we consider the sum of squared prediction error (SSPE) in this work.
Thus, the resulting loss function is
\begin{equation}\label{eq:loss}
  \begin{split}
\CV(\y_T, \y_V, \bzeta)= \sum_{i=1}^{n_V}(\widehat \y_{V, \bzeta[i]} - \y_{V[i]})^2,
\end{split}
\end{equation}
where $[i]$ indicates the $i$-th element of the vector. 
The CV estimate $\widehat \bzeta_\text{CV}$ is found by minimizing $\CV(\cdot)$ with respect to~$\bzeta$.

\spacing{.98}
This CV estimation is also known as \emph{hold-out validation} and differs from $k$-fold and leave-one-out CV in that only one division into training and validation data is considered~\citep{Arlo:Celi:10}.
In this study we rely on hold-out validation anticipating the target dataset sizes to be on the order of $10^6$--$10^7$ observations.
Such datasets are often sub-sampled before the GP model is fitted, \eg, the $5\times 10^6$ observations considered in Section~\ref{sec:data} are a random subset of more than $2.8 \times 10^7$ observations~\citep{Finl:Datt:etal:19}.
An alternative approach to handle such large datasets is to minimize~$\CV(\cdot)$ from~\eqref{eq:loss} by stochastic gradient-descent optimization, which considers only a random sample of the data at each iteration~\citep{rude:16}. 
In both cases it is appropriate to focus on hold-out validation from a larger sample of the data rather than using $k$-fold CV. 
We note that, if the dataset is small enough, it is straight forward to turn our parallel CV parameter estimation method into parallel $k$-fold~CV, but this is not developed in this work. 

Both the CV and ML based estimation of covariance parameters have been studied from a theoretical perspective \citep{Stei:90} and using simulation studies \citep{Sund:etal:01}.
An important distinction is whether the covariance model is misspecified, where misspecified means that the covariance function of the data cannot be represented by a parameter configuration of the fitted covariance model.
For the case where the model is correctly specified it is known that CV has lager asymptotic variance than ML for Brownian motion \citep{Stei:90}.
Conversely, in the misspecified case, CV can lead to smaller squared prediction errors \citep{Bach:18,Bach:13}. 
For most applications the true covariance model is not known, and hence, a misspecified covariance model is likely; in this practical situation CV is an attractive alternative to~ML.
Moreover, the simulation results from Section~\ref{sec:simMLE} suggest that even with a correctly specified covariance function CV inference can provide competitive predictions.  

\vspace*{-2mm}
\subsection{Parallel implementation}\label{sec:parallel}
To introduce a parallel version of~$\CV(\cdot)$ in~\eqref{eq:loss} we first consider the case where the computations are performed on $N=2$ central processing units (CPUs), \ie, at most two computations are performed in parallel. 
To that end, consider a rectangular spatial domain~$D$ divided into two disjoint rectangles $D_1$ and~$D_2$.
Let $\y^i_T$ and $\y^i_V$ denote the training and validation data vectors from subset~$D_i$.
Then an approximate version of $\CV(\cdot)$~is
\begin{equation}\label{eq:approx1}
\CV(\y_T, \y_V, \bzeta)\approx \CV\left(\y^1_T, \y^1_V, \bzeta\right) + \CV\left(\y^2_T, \y^2_V, \bzeta\right).
\end{equation}
Here $\CV\left(\y^1_T, \y^1_V, \bzeta\right)$ and $\CV\left(\y^2_T, \y^2_V, \bzeta\right)$ can be evaluated in parallel and the scalar results are added.

\begin{figure*}
\centering
\includegraphics[width=\textwidth]{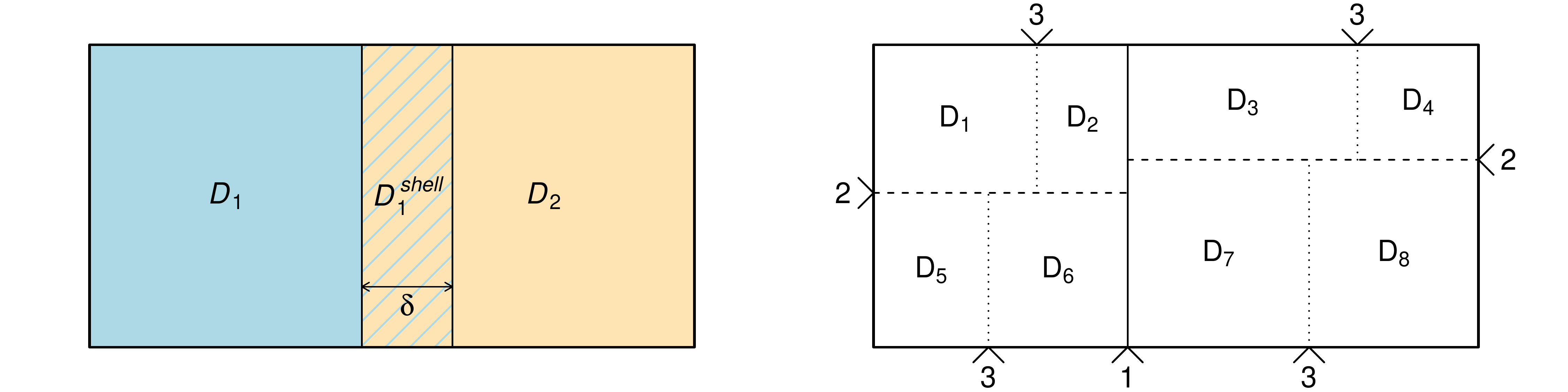}
\caption{Left: The spatial domain $D$ (entire rectangle) is divided into $N=2$ disjoint sub-domains $D_1$ (blue) and $D_2$ (orange).
The prediction of the validation data in $D_1$ is based on the training data in $D_1$ (blue) and its shell~$D_1^\text{shell}$ (cross-hatched).
Right: Recursive division of the spatial domain~$D$ (entire rectangle) into $N=8$ sub-domains $D_1, \dots, D_8$.
The numbers around~$D$ indicate the steps of the recursion and the solid, dashed, and dotted lines the corresponding splits
}
\label{fig:split}
\end{figure*}

\spacing{1}
Clearly, the approximation in~\eqref{eq:approx1} may be inaccurate, because the prediction of $\y^1_V$ lacks the information of the training data $\y^2_T$ and vice-versa.
To improve the approximation we assume that close observations are more relevant for the prediction than more distant ones.
Thus, the observations in $\y^2_T$ with a large potential to improve the predictions of $\y^1_V$ lie near the boundary of~$D_1$. 
We exploit this to improve the approximation in~\eqref{eq:approx1} as follows. 
Let $D_1^\text{shell}\subset D_2$ denote a \emph{shell} of~$D_1$, which is defined through the width~$\delta \geq 0$ as illustrated in Fig.~\ref{fig:split} (left).
Furthermore, let $\widetilde \y^1_T$ denote training data in $D_1\cup D_1^\text{shell}$ and construct $\widetilde \y_T^2$ similarly.
Then the improved approximation can be written as
\begin{equation}\label{eq:approx2}
  \begin{split}
    \CV(\y_T, &\y_V, \bzeta)\approx\,\, \widetilde \CV(\y_T, \y_V, \bzeta)\\
    = &\,\,\CV\left(\widetilde\y^1_T, \y^1_V, \bzeta \right) + \CV\left(\widetilde\y^2_T, \y^2_V, \bzeta \right). \hspace*{-1cm}
  \end{split}
\end{equation}

The shell width~$\delta$ controls the approximation accuracy and the computational workload.
That is, a large $\delta$ increases the number of observations in $\widetilde \y^i_T$, and hence, the prediction accuracy of~$\y^i_V$ as well as the size of the linear system to be solved. 
The relevant question is how small can~$\delta$ be relative to~$\bzeta$ and the number of observations in order to keep the approximation error below a certain bound.
A simulation study investigating this question is given in Section~\ref{sec:simBoundary}.

Our parallel CV method generalizes this concept to $N$ subsets of~$D$.
The domain~$D$ is divided into the subsets $D_1,\dots, D_N$ and $\widetilde \y^i_T$ and $\y^i_V$ are constructed accordingly.
Then the approximation is 
\begin{equation}\label{eq:approx3}
    \begin{split}
      \CV(\y_T, \y_V, \bzeta)\approx&\,\, \widetilde \CV(\y_T, \y_V, \bzeta)\\
      =&\,\,\sum_{i  =1}^N \CV(\widetilde \y^i_T, \y^i_V, \bzeta). 
\end{split}
\end{equation}

A pseudo code version of the parallel evaluation of~$\widetilde \CV(\cdot)$ is given by Algorithm~\ref{alg:pcv}. 
We see that each CPU only accesses one particular subset of the data at a time, which can be coordinated with a parallel file storage system.
Ideally with $N$~CPUs, each CPU can exclusively process the data of one subset.
This has the advantage that multiple sequential evaluations with different $\bzeta$ parameters can be done while dividing (line~2 and~3) and reading (line~5) the data have to be done once.
Another feature of the algorithm is that the communication among the CPUs is limited to receiving the~$\bzeta$ to be evaluated (line~6) and sending the local SSPEs to one CPU (line~9) for gathering.
Because $\bzeta$ is low dimensional and the SSPE is a scalar the amount of required communication is negligible.
Thus, the algorithm has good properties to work efficiently on common HPC infrastructures.

\algblock{ParFor}{EndParFor}
\algnewcommand\algorithmicparfor{\textbf{parallel\ for}}
\algnewcommand\algorithmicpardo{\textbf{do}}
\algnewcommand\algorithmicendparfor{\textbf{end\ parallel\ for}}
\algrenewtext{ParFor}[1]{\algorithmicparfor\ #1\ \algorithmicpardo}
\algrenewtext{EndParFor}{\algorithmicendparfor}
\begin{algorithm}
  \caption{Parallel evaluation of $\widetilde \CV(\y_T, \y_V, \bzeta)$}\label{alg:pcv}
  \vspace*{.2cm}
  \hspace*{\algorithmicindent} \textbf{Input:} $\y_T$, $\y_V$, $\bzeta$\\
    \hspace*{\algorithmicindent} \textbf{Output:} $\widetilde \CV(\y_T, \y_V, \bzeta)$
    \begin{algorithmic}[1]
      \Procedure{}{}
      \State Create $\widetilde \y^1_T, \dots,\widetilde \y^N_T $ from $\y_T$.
      \State Create $\widetilde \y^1_V, \dots,\widetilde \y^N_V $ from $\y_V$.
      \ParFor{$i = 1, \dots, N$}
      \State Read $\widetilde \y^i_T$ and $\y^i_V$ in memory.
      \State Receive $\bzeta$. 
      \State Compute local SSPE $z_i= \CV(\widetilde \y^i_T, \y^i_V, \bzeta)$.
      \EndParFor
      \State Combine to global SSPE $z=\sum_{i=1}^N z_i$. 
      \State \textbf{return} $z$
    \EndProcedure
  \end{algorithmic}
\end{algorithm}

\spacing{1.02}
The number of subsets and $\delta$ controlling the size of the shells $D_i^\text{shell}$ determine the computational cost of~$\widetilde \CV(\cdot)$.
The computational cost of~$\CV(\cdot)$ is dominated by the kriging prediction in~\eqref{eq:krig}, and hence, is of order~$\cO(n^3)$. 
Let~$k$ be the number of observations in the largest training subset~$\widetilde \y^i_T$.
Then the computational cost of~$\widetilde \CV(\cdot)$ in~\eqref{eq:approx3} is~$\cO(Nk^3)$.
If $N$~CPUs are used in parallel, the computation cost per CPU is at most~$\cO(k^3)$.
Thus, for large~$N$ and small $\delta$ the computational cost of~$\widetilde \CV(\cdot)$ is much smaller compared to a global evaluation of~$\CV(\cdot)$.
Details of the scaling properties of the parallel CV method are illustrated in Section~\ref{sec:scaling}.

\subsection{Division of the data into subsets}\label{sec:subsets}
To achieve a balanced workload for the parallel evaluation of~$\widetilde \CV(\cdot)$ from \eqref{eq:approx3} it is essential that all~$\widetilde \y_T^i$ consist of a similar amount of data.
We assume $\y$ to be contained in a rectangular domain~$D$ and use a simple recursive approach to divide it into the sub-domains $D_1,\dots, D_{2^q}$, $q \in \mathbb{N}$, which are in turn used to construct $\widetilde \y_T^i$ and~$\widetilde \y_V^i$.
The recursive step consists of dividing a rectangle into two rectangles such that both rectangles together with their shells contain a similar amount of data.
The division lines alternate between parallel to the $x$-axis and parallel to the $y$-axis for each step of the recursion.
For $\delta=0$ the recursive division leads to subsets with similar amounts of data.
However, for $\delta > 0$ and four or more subsets, the subsets can exhibit substantial variability in the amount of data.
Clearly, better division strategies can be found for that situation, and this is a topic for further investigation. 

We illustrate the division in Fig.~\ref{fig:split} (right), where $D$ is divided recursively into $N=8$ sub-domains.
In the first step, $D$ is divided along the solid line labeled by~$1$.
In the second step, the two resulting subsets are divided along the dashed lines labeled by~$2$.
Finally, in the third step, the four resulting subsets are divided along the dotted lines labeled by~$3$.

\section{Simulation studies}
For the remainder of the manuscript we assume an exponential covariance model.
The corresponding covariance function is $c\left(s_1, s_2, \theta\right)=\exp(-||s_1-s_2||/\theta)$, where $s_1, s_2 \in D$ are spatial locations, $|| \cdot||$ denotes the Euclidean norm, and $\theta >0$ is the spatial range.
The matrix $\bSigma(\btheta)$ from Section~\ref{sec:model} is constructed as
$\{c(s_i, s_j, \theta)\}_{i,j=1,\dots, n}$.
With that the parameters of the likelihood in~\eqref{eq:lik} are $\bxi=(\sigma^2,\tau, \theta)\T$, and we focus on $\bzeta=(\lambda, \theta)\T$ with $\lambda=\tau/\sigma^2$ as those are the relevant parameters for prediction.

\subsection{Comparison of CV and ML estimation}\label{sec:simMLE}
\begin{figure*}[]
\centering
\includegraphics[width=\textwidth]{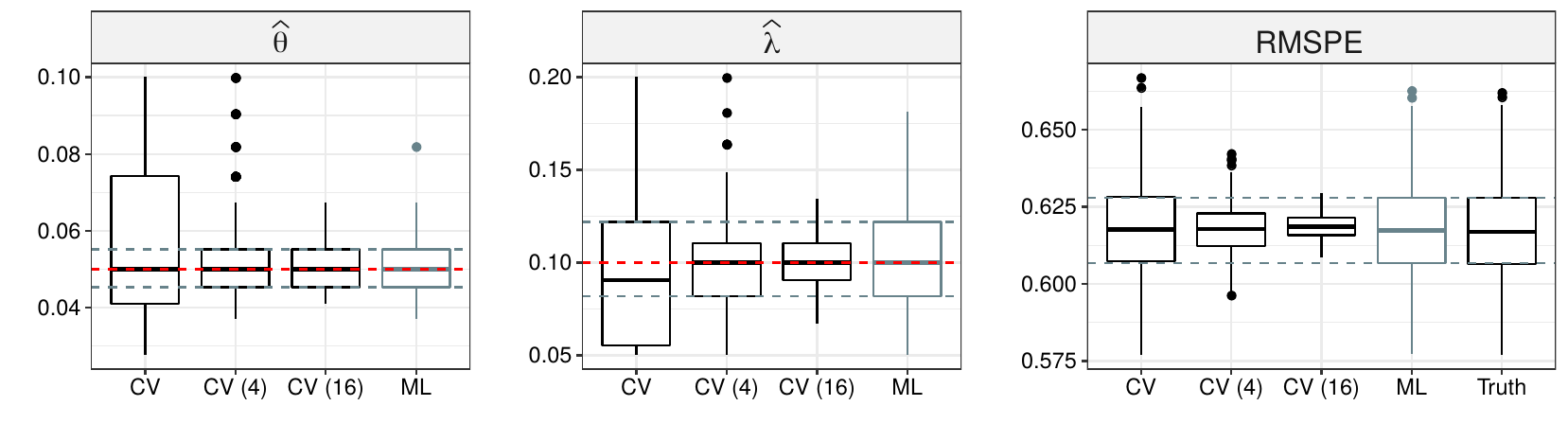}
\vspace*{-6mm}
\caption{Results of the simulation study comparing the CV and ML based covariance parameter estimates.
The $x$-axes denote the estimation methods, and \mbox{'(4)'} and \mbox{'(16)'} indicate that $4$ and $16$ replicates of the simulated data with $n_T=\numprint{4000}$ and $\numprint{16000}$ training data are used, respectively. 
The $y$-axes indicate the spatial range~$\theta$ (left panel), the noise-to-signal ratio~$\lambda$ (middle panel), and the RMSPE at the hold-out test locations (right panel).
The red dashed lines in the left and middle panel show the true $\theta$ and $\lambda$~value, respectively
}
\label{fig:simEst}
\end{figure*}

We assess the performance of CV and ML based covariance parameter estimation with a simulation study.
To this end, consider the spatial domain $D=[0,1] \times [0,1]$ and fix $\sigma^2=1$, $\theta=0.05$, and~$\tau=0.1$.
Then $400$ realizations of the GP with $\numprint{3000}$ spatial locations each are generated.
The following considerations apply to each of those $400$ realizations: 
The $\numprint{3000}$ spatial locations are sampled according to a Latin Hypercube sampling design from the R~package \emph{lhs}~\citep{R:lhs}.
We estimate $\bzeta=(\lambda, \theta)\T$ using CV and ML under the assumption that $\sigma^2=1$ is known, \ie, $\lambda=\tau$. 
More specifically, the $\numprint{3000}$ samples are randomly divided into $n_T=\numprint{1000}$ training, $n_V=\numprint{1000}$ validation, and $n_P=\numprint{1000}$ test data.
Then grid-search optimization is used to estimate the parameters, where the evaluated parameter grid consists of all $225$ pairwise combinations of $15$ parameters in $[\theta/2, 2\theta]$ and $15$ parameters in $[\lambda/2, 2\lambda]$.
The training and validation data of each realization are used to evaluate $\CV(\cdot)$ in~\eqref{eq:loss} and the parameter configuration leading to the smallest~$\CV(\cdot)$ is~$\widehat\bzeta_\text{CV}$.
Similarly, $\widehat\bzeta_\text{ML}$ is found by evaluating~\eqref{eq:lik} using the training data.
The prediction accuracy at the hold-out test locations is measured by the root-mean-square prediction error (RMSPE) for both the CV and ML estimates.
Finally, we study how the CV estimates change when $4$ and $16$ replicates of the simulated data are available.
Studying these replicates allows us to mimic larger datasets with $n_T=\numprint{4000}$ and $\numprint{16000}$ training data, respectively, while keeping the computational costs at a moderate level.

The left and middle panel of Fig.~\ref{fig:simEst} show boxplots of~$\widehat \theta$ and~$\widehat \lambda$, respectively.
All median estimates are near the true values, and the CV based estimates show larger variability around the true values than the ML estimates.
However, when CV is used for the datasets with $4$ and $16$ replicates, the variability around the true values is reduced.
The right panel of Fig.~\ref{fig:simEst} shows boxplots of the RMSPEs of the fitted models at the hold-out test locations.
Here both CV and ML show a similar distribution and both are close to the reference distribution, which is obtained using the true $\bzeta$ values for the prediction (rightmost boxplot). 
The variability of the RMSPEs is reduced when the CV estimation is applied to the datasets with $4$ and $16$ replicates.

\subsection{Choice of the shell width}\label{sec:simBoundary}
\begin{figure*}[]
\centering
\includegraphics[width=\textwidth]{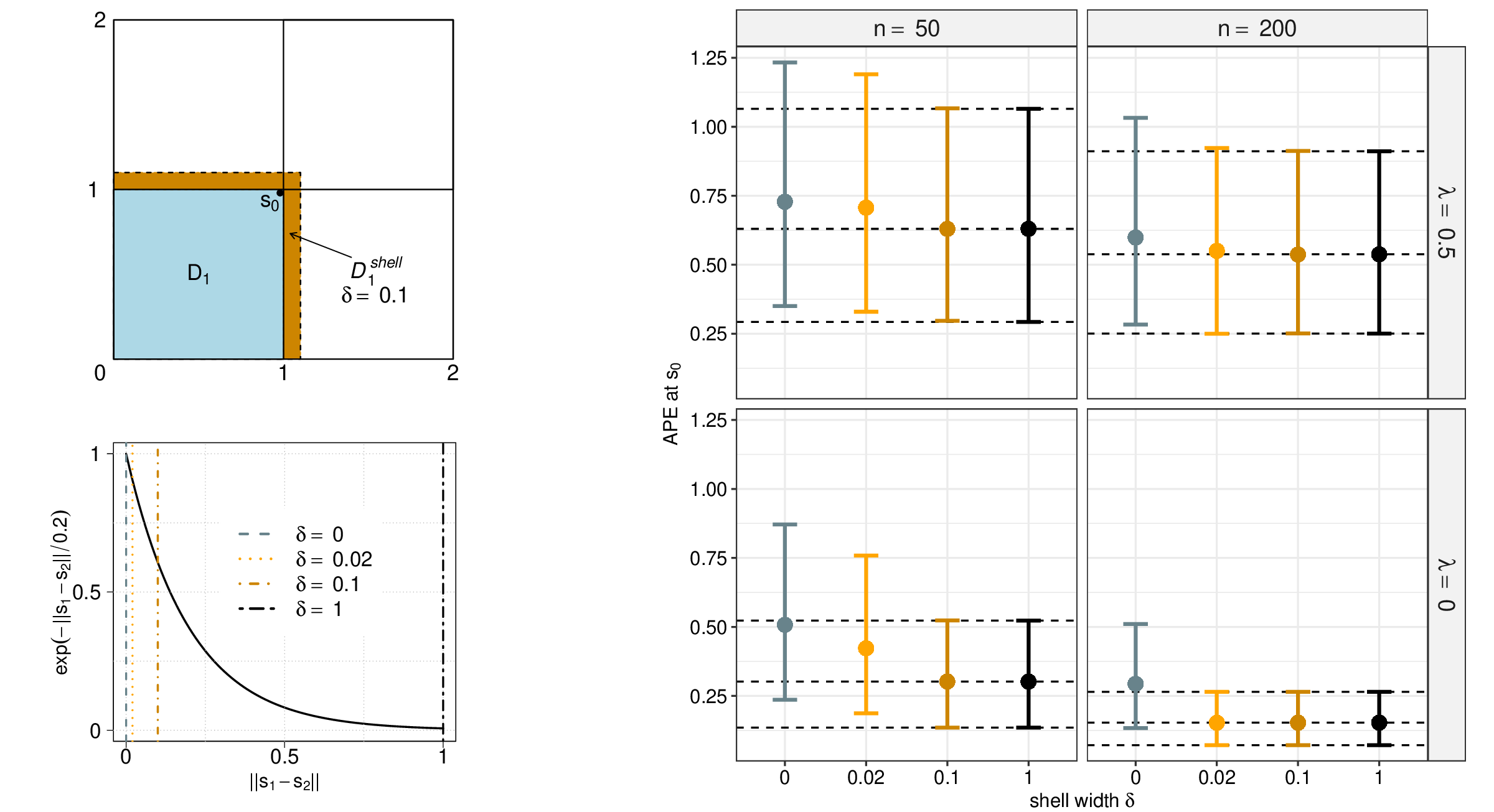}
\caption{Top left: The $[0,2] \times [0,2]$ square shows the spatial domain $D$ of the simulation study.
The GP is predicted at $s_0\in D_1$ given the data in $D_1 \cup  D_1^\text{shell} \setminus \{s_0\}$.
The prediction error is assessed for different shell widths~$\delta$.
Bottom left: The exponential covariance function with range $\theta=0.2$ and the considered shell widths~$\delta$ are shown.
Right: The median (dots) and quartiles (error bars) of the APEs at~$s_0$ ($y$-axes) are shown for the several~$\delta$ ($x$-axes).
The four panels show different scenarios, which vary in the number of simulated data~$n$ and the noise-to-signal ratio~$\lambda$.
Note that $\delta=1$ corresponds to using all simulated data in $D\setminus \{s_0\}$ for the prediction at~$s_0$ and serves as a reference (dashed lines)
}  
\label{fig:simBoundary}
\end{figure*}

The shell width~$\delta$ affects the prediction accuracy at the validation locations, and hence, the approximation accuracy of~$\widetilde \CV(\cdot)$ in~\eqref{eq:approx3}. 
The spatial locations with the largest potential to get sub-optimal predictions due to a small~$\delta$ are located near the corners of the subsets.
Conversely, if the predictions at those corner locations are good, the approximation error of~$\widetilde \CV$ is small.
While this consideration is useful to obtain a conservative estimate of the require shell width~$\delta$,
we have to keep in mind that the number of corner locations is small relative to the number of observations in the entire subset.
Thus, $\widetilde \CV(\cdot)$ can still be an accurate approximation even if the predictions of the corner locations are sub-optimal.

The following simulation setup is designed to assess how the prediction accuracy at a corner location depends on the shell width~$\delta$, the number of training data~$n$ relative to $\theta$, and~$\lambda$.
Let $D=[0,2] \times [0,2]$ be a spatial domain and $D_1=[0,1] \times [0,1]$ a subset thereof containing the spatial test location $s_0\in D_1$ near $(1,1)$ as shown in Fig.~\ref{fig:simBoundary} (top left).
Then we quantify the prediction accuracy at~$s_0$ given the data in $D_1\cup D_1^\text{shell}/\{s_0\}$ by the following procedure:  
First, sample $n$~spatial locations in~$D$ according to a space-filling sampling design.
Second, simulate the values of the GP defined in Section~\ref{sec:model} at the $n$ sampled locations and~$s_0$.
Third, predict $y_0=Y(s_0)$ based on the data with spatial location in $D_1 \cup  D_1^\text{shell} \setminus \{s_0\}$ using the true parameters~$\bzeta$.
Finally, quantify the prediction error at~$s_0$ as the absolute prediction error (APE) $|y_0 - \widehat y_0|$.

Fig.~\ref{fig:simBoundary} (right) shows the distribution of the APEs at~$s_0$ for a fixed spatial range~$\theta=0.2$ and
varying values for $\delta$, $n$, and~$\lambda$. 
The dots represent the median and the vertical lines the $25\%$ and the $75\%$~quantiles of the APEs from $\numprint{4000}$ simulations. 
Note that $\delta=1$ corresponds to using all simulated data in~$D$ and serves as a reference.
We see that for $n=50$ ($200$) a shell width of $\delta=0.1$ ($0.02$) is sufficient to obtain predictions that are comparable to using all data in~$D\setminus \{s_0\}$.
Note that $\delta=0.1$ is still much smaller than the effective spatial range of the covariance function (Fig~\ref{fig:simBoundary}, bottom left),
and this is a specific illustration of the screening effect~\citep{Stei:02}.
Surprisingly, the value of the noise-to-signal ration~$\lambda$ affects the APEs in the same way for all~$\delta$, and hence, is not relevant for the choice of~$\delta$.

\section{Data illustration}\label{sec:data}
\subsection{Dataset}
\begin{figure*}
\centering 
\includegraphics[width=\textwidth]{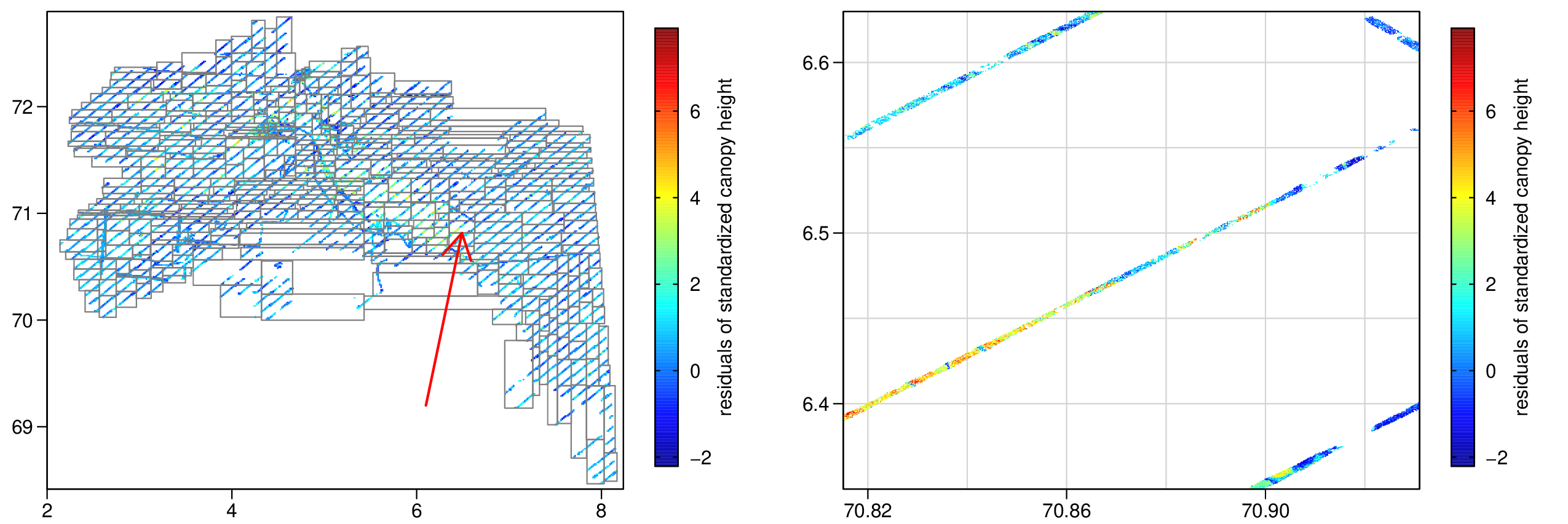}
\caption{Left: Map of the $5 \times 10^6$ residuals derived from the standardized LiDAR canopy height data from Alaska.
The gray lines indicate the division into $512$ rectangular subsets.
Right: An enlargement of the rectangle indicated by the red arrow on the left panel
}
\label{fig:data}
\end{figure*}

We consider a spatial dataset with $5\times 10^6$ spatial observations consisting of airborne LiDAR canopy height measurements taken in Alaska in 2014.
This dataset was previously used to study spatial models~\citep{Finl:Datt:etal:19,Tayl:etal:18}, and we refer the reader to those publications for more background on the dataset and its creation.
Due to the measurement process, the observations are available along strips as shown in Fig.~\ref{fig:data}.
Of interest, however, are a high-resolution map of the canopy height and measures of uncertainty for the entire spatial domain.  
Both can be obtained using a GP model, and we take this as motivation to study the parallel CV method using this dataset. 

In addition to the canopy height measurements and their spatial locations we have access to two covariates: records of forest fires and forest canopy sparseness.
To remove potential dependence of the canopy height on these covariates, we first fit a linear model with an intercept, both covariates, and the interaction of longitude and latitude to the standardized canopy heights.
The residuals of that model are shown in Fig.~\ref{fig:data} and we consider them as the~``$\y$'' from the zero-mean GP described in Section~\ref{sec:method}.

\expandafter\def\expandafter\UrlBreaks\expandafter{\UrlBreaks
  \do\a\do\b\do\c\do\d\do\e\do\f\do\g\do\h\do\i\do\j%
  \do\k\do\l\do\m\do\n\do\o\do\p\do\q\do\r\do\s\do\t%
  \do\u\do\v\do\w\do\x\do\y\do\z\do\A\do\B\do\C\do\D%
  \do\E\do\F\do\G\do\H\do\I\do\J\do\K\do\L\do\M\do\N%
  \do\O\do\P\do\Q\do\R\do\S\do\T\do\U\do\V\do\W\do\X%
  \do\Y\do\Z}

\spacing{1.}
\subsection{Computing environments}
All computationally intensive tasks are performed on \emph{Google Cloud}\footnote{\url{https://cloud.google.com}}\hspace{-1pt}.
Depending on the memory requirements of the computation either up to $514$ n1-\mbox{standard-1} nodes with one Intel Xeon CPU at $2.30$\,GHz and $3.75$\,GB memory or
up to $258$ n1-\mbox{highmem-2} nodes with $2$~Intel Xeon CPU at $2.30$\,GHz and $13$\,GB memory are used. 
We set up the nodes as an \emph{elastic Slurm cluster}\footnote{\url{https://slurm.schedmd.com/elastic_computing.html}} 
with a \emph{CentOS~7}\footnote{\url{https://www.centos.org}} Linux operating system and the statistical software~\citet{R}.
Parallel computations are performed using \emph{OpenMPI}\footnote{\url{https://www.open-mpi.org}} directives, which are formulated using the R~package \emph{pbdMPI}~\citep{R:pbdMPI}.
OpenMP multi-threading is disabled.
Computations using fewer than $80$~CPUs are performed on a university owned node with $80$~Intel Xeon CPUs at $2$\,GHz and a total of $2$\,TB memory. 
The code and data to reproduce our results are available at:
\url{https://github.com/florafauna/parallelCVsupplementaryMaterial}.

\subsection{Scaling experiments}\label{sec:scaling}
\subsubsection{Strong scaling}\label{sec:strong}
\begin{figure*}
\centering
\includegraphics[width=\textwidth]{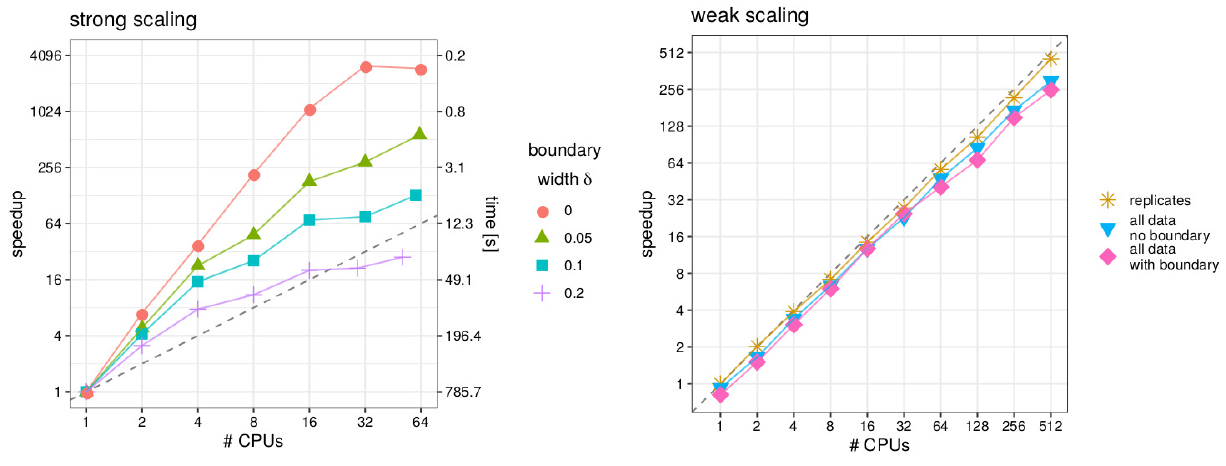}
\caption{Left: Results of the strong scaling experiment shown as the speedup relative to using one CPU \mbox{(y-axis)} for different number of CPUs \mbox{(x-axis)} and four shell widths~$\delta$.
Right: Results of the weak scaling experiment depicted as the speedup relative to the fastest subset \mbox{(y-axis)} and the number of used CPUs \mbox{(x-axis)}.
The dashed lines indicate perfect scaling
}
\label{fig:scaling}
\end{figure*}

Strong scaling describes the parallel computing speedup relative to the number of CPUs for a fixed problem size.
To measure it we consider a subset of the data with $\numprint{20000}$ residuals and rescale the corresponding spatial domain to $D=[0,1]\times [0,1]$.
The data is randomly divided into $n_T=\numprint{18000}$ training and $n_V=\numprint{2000}$ validation data.
Using the recursive division described in Section~\ref{sec:subsets} with $0,\dots, 6$ recursive steps we create seven datasets with $N=1,2,4,8,16,32$, and $64$ subsets. 
For the division we consider the shell widths $\delta = 0$, $0.05$, $0.1$, and~$0.2$, and create one series of subsets for each~$\delta$ leading to a total of $28$ datasets.
Note that $\delta=0.2$ is $1/5$ of the domain width and most applications would not require~$\delta$ this large.
The computation time of $\widetilde \CV(\cdot)$ from~\eqref{eq:approx3} is measured for each of these datasets, where the number of used CPUs is set to $N$, \ie, the number of subsets of the specific dataset.

The results of the scaling study are shown in Fig.~\ref{fig:scaling} (left).
One evaluation of~$\widetilde \CV(\cdot)$ using one CPU, \ie, $\widetilde \CV(\cdot)=\CV(\cdot)$, takes $786$ seconds.
When instead $\widetilde \CV(\cdot)$ is evaluated using multiple CPUs in parallel, the speedup is in most cases larger than the number of used CPUs.
This is expected, as the approximation $\widetilde \CV(\cdot)$ has a computational cost of $\cO(Nk^3)$, whereas $\CV(\cdot)$ has a cost of $\cO(n_T^3)$ (see Section~\ref{sec:parallel}). 
Moreover, the shell width~$\delta$ determines the amount of processed data and has a large impact on the scaling property.
For $\delta>0$ the recursive divisions lead to some degenerate subsets without validation data.
We ignored those subsets and this explains why some timing results in the figure are plotted against less CPUs than one would expect.  
For $\delta=0$ the evaluation time reaches a lower bound when using $32$ CPUs in parallel, which suggests that the execution of non-parallel parts of the code together with the parallel computing overhead take about $0.3$~seconds.

\spacing{1.04}
\subsubsection{Weak scaling}\label{sec:weak}
Weak scaling describes the parallel computing speedup relative to the number of CPUs when the problem size increases linearly with the number of CPUs.
To measure it we consider all $5\times 10^6$ residuals and randomly choose $n_T=\numprint{3999462}$ ($80\%$) training data and $n_V=\numprint{500346}$ ($10\%$) validation data.
We consider three ways to generate the subsets and their shells:
First, we divide the data into $512$ subsets without shells ($\delta=0$) as described in Section~\ref{sec:subsets}.
In that case the subsets have either $\numprint{9765}$ or $\numprint{9766}$ residuals with between $\numprint{7698}$ and $\numprint{7936}$ training data.
Second, we consider a shell width of $\delta=0.001$ and again divide the data into $512$ subsets.
The division leads to $20$~subsets with more than $\numprint{2000}$ residuals in their shells, and we reduce the number of residuals in those shells to $\numprint{2000}$ by random sampling.
The resulting $512$ subsets have between $\numprint{9966}$ and $\numprint{13049}$ residuals with between $\numprint{7863}$ and $\numprint{10478}$ training data.
Third, we construct a dataset consisting of $512$~replicates of one subset from the first case with $\delta=0$ above.
In that case, all subsets have $\numprint{9766}$ residuals with $\numprint{7845}$ training data, and hence, it mimics a perfect division of the dataset into subsets.
For the three series of subsets we measure the evaluation time of~$\widetilde \CV(\cdot)$ using different numbers of subsets and the corresponding number of~CPUs. 

Fig.~\ref{fig:scaling} (right) shows the parallel computing speedup for the three series of subsets.
Not surprisingly, the series consisting of replicates has the best and almost perfect scaling.
That is, the evaluation time of~$\CV(\cdot)$ using one subset and one CPU is similar to the evaluation of$~\widetilde \CV(\cdot)$ using all $512$ replicates and $512$ CPUs.
The other two series of subsets have a varying number of training data in each subset, and hence, the evaluation time of~$\CV(\cdot)$ varies for each subset.
Therefore, it is unclear against which subset the evaluation times should be compared in order to compute the speedup.
We decided to use the fastest subset as a reference, which is the most conservative approach.
With that the scaling is promising but less optimal than for the series consisting of replicates.
The deviation from perfect scaling is mostly due to the imperfect division of the data into subsets, which leads to unbalanced workloads among the~CPUs.

\subsection{Parameter estimation}
\begin{figure*}
\centering
\includegraphics[width=\textwidth]{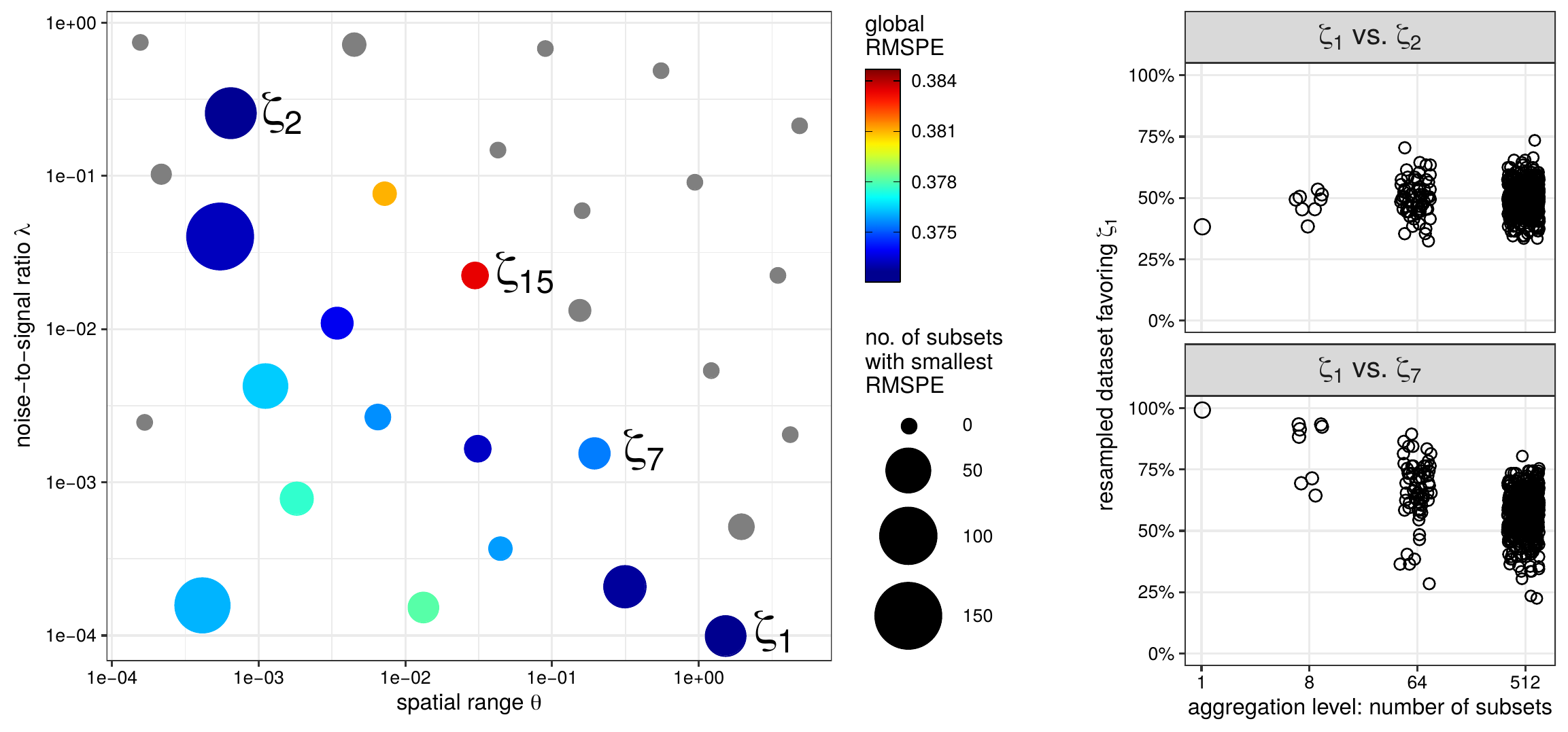} 
\caption{Left: Results of the parallel CV covariance parameter estimation for the $5\times 10^6$ residuals of the canopy height data. 
Each dot represents one evaluated parameter configuration $\bzeta_i$, where the~$\lambda$ and~$\theta$ components are shown on the $x$ and \mbox{$y$-axis}, respectively.
The colors indicate the global RMSPEs and gray points have a RMSPE greater than~$0.384$.
The labels $\bzeta_1$, $\bzeta_2$, $\bzeta_7$, and~$\bzeta_{15}$ denote the parameter configuration with the smallest, the $2$nd, $7$th, and $15$th smallest global RMSPEs, respectively.  
Dots sizes indicate the number of subsets for which the parameter configuration lead to the smallest local RMSPE.
Right: Assessment of the influence of the random sampling of the training and validation data.
The percentage of the $100$ resampled datasets favoring~$\bzeta_1$ (\mbox{$y$-axis}) is plotted against the aggregation level indicated by the number of subsets (\mbox{$x$-axis})
}
\label{fig:fit}
\end{figure*}

We used parallel CV to fit the GP model outlined in Section~\ref{sec:method} to the residuals of the canopy height dataset.
To that end, the $5\times 10^6$ residuals are randomly split into $80\%$ training, $10\%$ validation, and $10\%$ test data.
The data are divided into $512$ subsets using a shell width of $\delta=0.001$ as previously described in Section~\ref{sec:weak}.
Then $\widehat \bzeta_\text{CV}=(\widehat \lambda_\text{CV}, \widehat \theta_\text{CV})\T$ is found via grid-search optimization, where the grid consists of $30$ parameters $\bzeta_i$, $i=1,\dots,30$,
which are chosen based on a Latin Hypercube sampling design from the R~package \emph{lhs}~\citep{R:lhs}.
Then $\widetilde \CV(\cdot)$ of~\eqref{eq:approx3} is evaluated sequentially for all~$\bzeta_i$ using the training and validation data.
\emph{The computations are performed using $512$~CPUs in parallel and take $44$~minutes in total.}
Hence, one evaluation of~$\CV(\cdot)$ takes $1.4$ minutes on average and a total of $376$~CPU hours are used.

\spacing{.97}
From the results of $\widetilde \CV(\cdot)$ it is straight forward to compute the approximate global RMSPEs for all~$\bzeta_i$ as shown in Fig.~\ref{fig:fit} (left).
Here \emph{global} indicates that all subsets are used for the computations.
We reorder the $\bzeta_i$ according to their global RMSPE, \ie, $\bzeta_1=\widehat\bzeta_\text{CV}$ indicates the parameters with the smallest and $\bzeta_{30}$ the one with the largest global RMSPE. 
In figure we see that $\bzeta_1$ and $\bzeta_2$ lie in opposite corners of the parameter space, which could be due to an oversimplified covariance model.

As opposed to the global RMSPE, we can also investigate the local RMSPE for each subset to get insights into non-stationary features of the data.
The circle sizes in the Fig.~\ref{fig:fit} (left) indicate the number of subsets for which the $\bzeta_i$ leads to the smallest local RMSPE.
We see many large dots for small $\theta$~values, and hence, the small range dependency in the data seems to be important.
Moreover, $19$ of the $30$ $\bzeta_i$ lead to the smallest local RMSPE for at least one of the $512$ subsets, which indicates that a non-stationary extension of the model could be beneficial. 
To further investigate this conjecture the hold-out test data are predicted using the global $\widehat\bzeta_\text{CV}$ and using the best local parameter configuration for each subset.
The resulting RMSPEs for the test data using the global and local estimates are $0.372$ and $0.366$, respectively.
This suggests that the non-stationary version provides more accurate predictions.

\spacing{1.02}
Furthermore, we assess the influence of the random splitting into training, validation, and test data on the parameter estimates.
To that end, we resample that division $100$~times, select $\bzeta_1$, $\bzeta_2$, $\bzeta_7$, and~$\bzeta_{15}$, and evaluate $\widetilde \CV(\cdot)$ for each resampled dataset and those parameters.
Using $512$ CPUs in parallel the computation takes $9.1$ hours, which correspond to a total of $\numprint{4659}$ CPU hours.
A comparison of the resulting global RMSPEs reveals that the ordering of $\bzeta_1$ and $\bzeta_2$ is only reproduced for $39\%$ of the resampled datasets, and hence, it is not clear which of those parameters is preferred. 
For the remaining five pairwise comparisons the ordering is confirmed for all resampled datasets.
In addition, we can investigate the local behavior of the RMSPEs at subset level and for different spatial aggregations. 
Fig.~\ref{fig:fit} (right) shows the proportion of resampled datasets favoring $\bzeta_1$ against $\bzeta_2$ (top, $y$-axis) and $\bzeta_1$ against $\bzeta_7$ (bottom, $y$-axis) for different spatial aggregation levels ($x$-axis).
The bottom panel shows that some subsets favor $\bzeta_7$ at the aggregation levels with $512$ and $64$ subsets, whereas at the global level $\bzeta_1$ is favored for $100\%$ of the resampled datasets.
This is another indication that a non-stationary model could be advantageous.

\spacing{1.04}
\section{Conclusion}
\vspace*{4mm}
In this work we revisit two old ideas for handling large spatial datasets: domain decomposition or subsetting and out-of-sample CV for parameter estimation.
Based on numerical results we show that a modest overlap of the subsets (referred to as \emph{shells}) provides an accurate approximation to a global spatial analysis.
An explanation for this useful property is the screening effect for spatial prediction.
For example, with a moderate sample size ($n= 200$) in a squared domain and an exponential covariance function with a range of $10\%$ of the domain width, the shell width~$\delta$ can be set to as low as $1\%$ of the domain width while recovering accurate predictions.
Thus, under the kind of GP models typically used for environmental applications one can use subsets with little overlap and still expect predictions that are comparable to a global model.
The size of the shell regions depends on the correlation range of the process and the observation density but less so on the noise-to-signal ratio~$\lambda$.
In practice one can get a rough idea of the correlation range through a simple exploratory analysis using variograms to determine a good choice for~$\delta$. 

As expected, CV estimates of the covariance parameters show larger uncertainty compared to the ML estimates in our simulation study.
In addition, we note several important findings.
The RMSPEs of predictions at test locations using either the CV or ML estimates are nearly identical.
This suggests that spatial prediction is robust to the estimated covariance model and that the CV parameters are adequate for prediction.
We also find that the accuracy of the CV estimates improves substantially with replicated fields.
Many environmental datasets exhibit (pseudo) replicated fields in the form of nearly uncorrelated and stationary fields over time and so this feature can be exploited to improve CV parameter estimates.
Note that the design of our simulation study puts ML in its best light by fitting the correct covariance model.
It would be interesting to see what level of misspecification makes the ML and CV parameter estimates comparable in mean squared error. 

Based on shell sizes that give good approximations to global predictions, our parallel CV method is straightforward and shows nearly optimal weak scaling results.
For strong scaling we see more dependence on the shell size but even with a generous shell width~$\delta$ of $20\%$ the evaluation is $16$ times faster using $16$ CPUs compared to using one CPU.
In the case of a smaller, but still realistic shell widths, we see a speedup of more than $250$ for $32$ CPUs.
Moreover, we consider a scientifically relevant dataset with $5$~million spatial observations as a practical benchmark, and we are able to estimate the covariance parameters based on \emph{all} data using $512$ CPUs in parallel for $45$ minutes.
The good scaling properties are expected given the limited amount of communication among the CPUs and the reduced amount of input/output per CPU.
That being said, there is still room for improving the scaling properties.
For example, we rely on a simple, recursive division to generate subsets.
Finding a more sophisticated division strategy that leads to a more balanced workload among the CPUs is a topic for future research. 

To keep our numerical examples simple we have omitted a mean component from the spatial model.
However, typical spatial process models include a linear regression component and the extended model of~\eqref{eq:mod} is
 \[    \y \sim \mathcal{N} \left( \X \bbeta, \sigma^2 \bSigma + \tau \I  \right),
 \]
where $\X$ is a matrix of covariates and $\bbeta$ a vector of linear predictors. 
Often $\bbeta$ is found using generalized least squares (GLS) and the corresponding estimate is 
\begin{equation}
  \label{GLS}
  \widehat{\bbeta} = (\X^T \M^{-1}  \X)^{-1} \X^T \M^{-1} \y,
\end{equation}
where $\M =  \bSigma + \lambda \I $.
$\M$ involves the process covariance for all locations, and hence, this may seem problematic for our subsetting approach. 
But the following considerations can be used to formulate the GLS estimation in terms of the already solved prediction problem.
Recall that the spatial prediction of the residuals~$\w$ at observed locations is given by~$\H\w$, where $\H=\bSigma (\bSigma + \lambda \I)^{-1}$.
Simple matrix identities imply that $\M^{-1}=(1/\lambda)(\I - \H)$, and hence, $\M^{-1} \X$ from~\eqref{GLS} can be found by a spatial prediction operation on the columns of $\X$.
Because the parallel CV method is already efficient for finding approximate predicted values, there will be only a minimal overhead in finding approximate GLS estimates of~$\bbeta$ for a modest number of covariates. 
 
One potential limitation of our parallel CV method concerns spatial processes with long range correlations that induce dependence across a large fraction of the domain.
One strategy to handle this situation is to include a low dimensional set of basis functions in the mean component to adjust for large scale dependence.
This can often reduce the correlation scale and also simplify the dependence structure.
Moreover, the basis function parameters can be computed based on GLS as outlined above. 

The main focus of this work is  a practical route to approximate a global spatial analysis.
While we have been successful in establishing an accurate parallel CV method to this end, we also note that local predictions on subsets of the domain have value on their own.
One might expect large spatial datasets that span a heterogeneous environment to exhibit a non-stationarity covariance structure.
For this reason it is natural to consider local covariance models for subsets of the domain.
Information about parameters and prediction errors at subsets scale is an intermediate computation from our method and readily available.
Section~\ref{sec:data} illustrates some ways to use the local results to assess stationarity and draws the tentative conclusion that a non-stationary model is more appropriate.
The fitting and prediction with non-stationary models is still an active area of research and raises the questions of how to identify and incorporate changing covariance parameters over space.
It would be interesting to apply the parallel CV method to such a non-stationary model, and we believe that the computational benefits as well as the interpretation of the local results can support progress in this research area.

\spacing{1.022}

\begin{acknowledgements}
We thank Google and the Computational Information Systems Laboratory of the University Corporation for Atmospheric Research for the provided compute time on their high-performance computers. 
\end{acknowledgements}

%
%

\bibliographystyle{spbasic}      
\bibliography{parallelCV}   

\end{document}